\documentclass[prl,aps,twocolumn,amsmath,amssymb,floatfix,reprint,footinbib,showpacs,superscriptaddress]{revtex4-1}
\usepackage[dvipsnames]{xcolor}
\usepackage{amsmath}
\usepackage{amssymb}
\usepackage{amsfonts}
\usepackage{epsfig}
\usepackage{tikz}
\usepackage{dsfont}
\usepackage{url}
\usepackage{graphicx}
\usepackage[colorlinks]{hyperref}
\hypersetup{%
	plainpages=true,
	breaklinks=true,
	hypertexnames=false,
	pageanchor=true,
	colorlinks=true,
	linkcolor={blue},
	citecolor={red},
	urlcolor={blue},
	anchorcolor={black}
}
\usepackage{bbm}
\usepackage[draft,inline,nomargin,index,marginclue]{fixme}
\graphicspath{{.}{./figs/}}
\newcommand{\ket}[1]{{\vert #1 \rangle}}

\begin{document}
\title{Multi-emitter oscillating bound states in Waveguide QED}

\author{Sergi Terradas-Brians\'o}
\affiliation{Instituto de Nanociencia y Materiales de Arag\'on (INMA), CSIC-Universidad de Zaragoza, Zaragoza 50009, Spain}\affiliation{Departamento de Fisica de la Materia Condensada, Universidad de Zaragoza, 50009 Zaragoza, Spain}
\author{Carlos A. Gonz\'alez-Guti\'errez}
\affiliation{Instituto de Ciencias F\'isicas, Universidad Nacional Aut\'onoma de M\'exico, Cuernavaca 62210, Mexico}
\author{Iv\'an Huarte}
\author{David Zueco}
\author{Luis Martin-Moreno}\affiliation{Instituto de Nanociencia y Materiales de Arag\'on (INMA), CSIC-Universidad de Zaragoza, Zaragoza 50009, Spain}
\affiliation{Departamento de Fisica de la Materia Condensada, Universidad de Zaragoza, 50009 Zaragoza, Spain}

\date{\today}
\begin{abstract}
Waveguide quantum electrodynamics platforms have emerged as promising candidates for exploring and implementing non-Markovian quantum phenomena. 
In this work, we investigate the formation and dynamics of superpositions of bound states in a cavity array waveguide coupled to two spatially separated quantum emitters. 
By tuning the system parameters, we show that spontaneous emission can drive the system into non-local equilibrium states in which both photonic and emitter populations exhibit persistent oscillations.
These states arise from the coexistence of bound states embedded in the energy continuum and bound states outside it, leading to hybrid oscillatory modes.
We analytically derive the conditions required for the emergence of these states, numerically simulate their formation through spontaneous emission, and predict their long-time behaviour.
Our results demonstrate that such bound-state superpositions enable the generation of emitter-emitter interaction through free evolution, while supporting oscillatory breathing modes of the photon density between the emitters.
\end{abstract}

\maketitle
%
%

Waveguide quantum electrodynamics (WQED) has emerged in recent years as a powerful platform for the implementation and control of quantum optical systems \cite{Roy2017,Sheremet2023, Ding2025,vanDiePen2025}. 
Recent experimental advances have enabled the strong coupling of quantum emitters to photonic waveguides with finite bandwidth, leading to the realisation of atom-photon bound states in systems based on superconducting qubits and resonators.
These platforms provide a practical and highly controllable setting for investigating fundamental phenomena in quantum optics \cite{liu2017, PRX_Gassparinetti2022}.
In WQED, bound states correspond to entangled light-matter states in which the photonic wave function is spatially localised around the emitter position \cite{Calajo2016, Sanchez-Burillo2017, Calajo2019}. 
This localisation suppresses radiative decay into propagating modes, resulting in a finite probability for the excitation to remain trapped in the emitter-field system at long times. This effect is commonly referred to as fractional decay \cite{John1984, John1987}.
Conventionally, such bound states appear just outside the waveguide band when the emitter excitation energy approaches the band edge and are therefore known as bound states outside the continuum (BOCs).
However, a different class of bound states can exist within the energy continuum itself, as originally proposed by von Neumann and Wigner\cite{vNeumann1929}.  These bound states in the continuum (BICs) have since been identified in a wide range of physical systems \cite{StillingerPRA1975,NimrodPRL2003,Longhi2007,Longhi2020}; a comprehensive review can be found in Ref.~\cite{Hsu2016}.
In the context of WQED, BICs can arise either from interference effects in systems of spatially separated emitters \cite{Tufarelli2013,Gonzalez-Tudela2017,Kanu2020,Kanu2022,KanuArxiv2023,Barahona-Pascual2026} or from structured light-matter interactions involving giant emitters coupled to multiple points along a waveguide \cite{Soro2023}.
More recently, it has been shown that multiple BICs can coherently combine to form oscillating bound states. This phenomenon was first demonstrated for giant emitters coupled to continuous waveguides \cite{Guo2020a}, where it was established that three equally spaced coupling points constitute the minimal configuration required to observe oscillatory dynamics involving two BICs.
By contrast, when the oscillations occur between a BIC and a BOC, a giant emitter with only two coupling points is sufficient \cite{Zhang2023MagicCav}.
The emergence of oscillating bound states induced by giant emitters has also been explored beyond the rotating-wave approximation, extending into the ultrastrong coupling regime \cite{Terradas2022Ultrastrong}.

In this paper, we introduce a platform distinct from giant-atom implementations that supports the emergence of oscillating bound states.
Specifically, we consider two quantum emitters coupled to a photonic crystal waveguide and show that spontaneous emission from a single emitter can populate superpositions of bound states. These superposed states give rise to non-local oscillatory dynamics, with populations coherently exchanged between photonic and matter degrees of freedom.
We present a detailed analytical and numerical analysis of the system dynamics, identifying the conditions under which such oscillating bound states form during spontaneous emission.
By appropriately tuning the system parameters, we show that an initially excited emitter can relax into a steady state consisting of a superposition of up to three bound states.
Our results establish a new and experimentally accessible configuration for realizing oscillating bound states, thereby broadening the class of waveguide QED systems in which such phenomena can be observed and providing a new route towards generation of emitter-emitter interactions in waveguide QED platforms.
%
%

\begin{figure}[!t]
 \centering
 \includegraphics[width = \columnwidth]{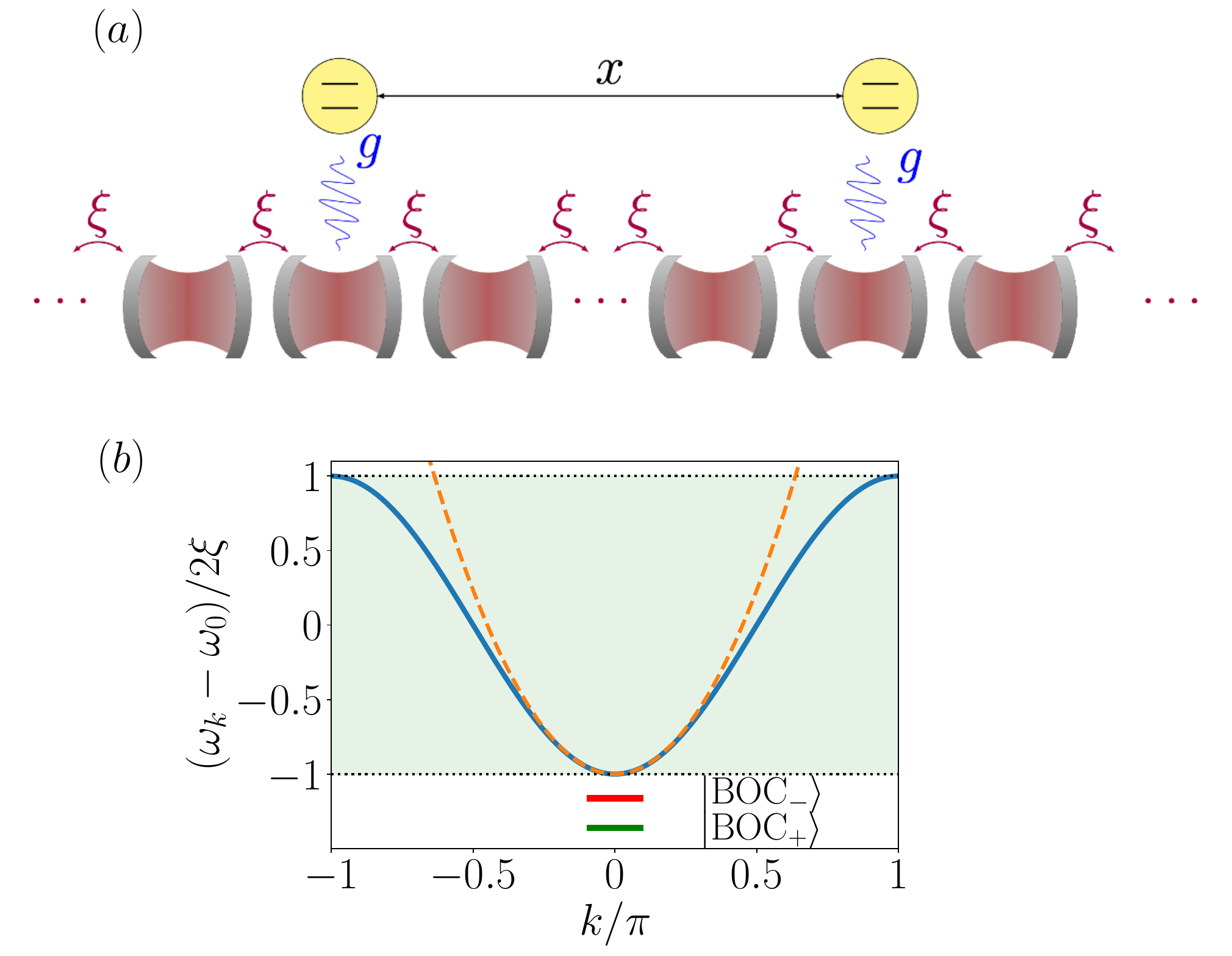}
 
 \caption{(a) Schematic representation of a waveguide composed of an infinite cavity array coupled to two quantum emitters.  The intercavity hopping amplitude is denoted by $\xi$, and the emitter-cavity coupling strength by $g$. Each cavity has a resonant frequency $\omega_{0}$, while the excitation energy of each emitter is $\Delta$. The two emitters are coupled to cavities separated by $x$ intermediate cavities.
(b) Dispersion relation of the cavity-array waveguide (solid blue line). The dashed orange line shows the dispersion relation of a continuous waveguide with similar low-energy behaviour.
The cavity-array band is bounded by dotted black lines and highlighted by the green shaded region.
Below the lower band edge, we indicate the energies of two hypothetical bound states outside the continuum arising from the coupling of the two emitters shown in (a).}
\label{fig:Scheme}
\end{figure}

We consider a system of two emitters, modelled as two-level systems, coupled to a waveguide, as schematically shown in Fig.~\ref{fig:Scheme}(a).
The waveguide is modelled as an array of coupled cavities. This description may correspond either to an actual cavity-array implementation or to a discretised representation of a continuous waveguide. In the latter case, the discretised model accurately reproduces the low-energy dynamics provided that the discretisation mesh (the intercavity distance) is much smaller than the relevant length scales of the system, such as the optical wavelength or the localisation length. By contrast, the dynamics near the upper band edge of the cavity array differ from those of a continuous waveguide, unless the latter also exhibits a spectral gap.
The $j$th emitter has a transition frequency $\Delta_{j}$ and couples to the cavity located at position $n_{j}$. Its coupling strength to cavity mode $k$ is denoted by $g_{j,k}$.
We assume that the light-matter coupling is sufficiently weak, such that the rotating-wave approximation is valid. Under this assumption, the dynamics of the system is governed by a spin-boson type Hamiltonian
\begin{equation}
    \label{eq:ArbH}
    H  \! =   \!\! \sum_k \omega_k a^\dagger_k a_k +  \!\! \sum_j \Delta_j \sigma^{+}_j \sigma^{-}_j 
     \!+   \!\!\sum_{j,k}  \! \left(g^{*}_{j,k} a^\dagger_k \sigma^{-}_j + \mathrm{h.c.}  \! \right) \!,
\end{equation}
where $a_k$ ($a^\dagger_k$) denotes the annihilation (creation) operator of the $k$th bosonic mode of the waveguide.
The operators $\sigma^{+}_j$ and $\sigma^{-}_j$ respectively create and annihilate an excitation in the $j$th emitter.
The waveguide dispersion relation is given by $\omega_k = \omega_0 - 2\xi \cos(k)$.
Figure~\ref{fig:Scheme}(b) shows both the dispersion relation $\omega_k$ of the coupled-cavity array and that of a hypothetical continuous waveguide exhibiting the same low-energy dynamics.
The figure also indicates the energies of the BOCs that may arise in the configuration sketched in Fig.~\ref{fig:Scheme}(a).
%
We fix the intercavity hopping to $\xi = 1$, thereby setting the energy scale, and measure all frequencies relative to $\omega_0$, which can be gauged away without affecting the dynamics.
Throughout this work, we consider two identical emitters with transition frequencies $\Delta_1 = \Delta_2 = \Delta$, coupled to the waveguide with equal strength $g_{j,k} = g e^{ik n_j}/\sqrt{N_k}$, where $N_k$ is the total number of waveguide modes.
We focus on the spontaneous emission dynamics initiated by a single excited emitter. Since the Hamiltonian conserves the total number of excitations, the analysis can be restricted to the single-excitation manifold.

To compute the spontaneous emission dynamics, we employ the resolvent operator method \cite{Cohen-AtPhInt-Ch3}, a powerful technique for obtaining the exact time evolution of states in the spin-boson model considered here.
The resolvent operator associated with a Hamiltonian $H$ is defined as $G(E) = (E - H)^{-1}$. By projecting $G(E)$ onto a set of states ${\ket{\alpha}}$, one obtains the corresponding Green functions, which can be Fourier transformed to yield the time-dependent state amplitudes $c_\alpha(t)$.
This approach also allows us to explicitly identify the contributions of bound states, which originate from the poles of $G(E)$.


Our primary objective is to describe the spontaneous emission dynamics. 
To this end, we work in the basis of maximally entangled states $\ket{\pm} = (\ket{L} \pm \ket{R})/\sqrt{2}$, where $\ket{L}$ ($\ket{R}$) denotes a state with a single excitation in the left (right) emitter.
The $\ket{\pm}$ basis is particularly convenient because it diagonalises the resolvent operator when projected onto the single-emitter excitation subspace.  As a consequence, the time evolution of the $\ket{\pm}$ states is decoupled, such that $\langle \mp | e^{-iHt} | \pm \rangle = 0$ at all times \cite{Gonzalez-Tudela2017,Soro2023}.
As a result, the amplitude evolution of an arbitrary emitter state $\ket{e}$ within the single-excitation manifold can be written as \cite{Sanchez-Burillo2017,Berman2010,Berman2010a,Gonzalez-Tudela2017}
\begin{eqnarray}
\label{eq:Amplitude}
c_e(t) = \frac{1}{2\pi i} \sum_{\alpha = \pm} \int \frac{\langle \alpha | e \rangle e^{-iEt} dE}{E - \varepsilon_\alpha - \Sigma_\alpha(E)} ,
\end{eqnarray}
where $\varepsilon_\pm = \langle \pm | H_0 | \pm \rangle$ denotes the energy expectation value of the states $\ket{\pm}$ with respect to the non-interacting part of the Hamiltonian~\eqref{eq:ArbH}.
The self-energy terms are given by $\Sigma_\pm(E) = \frac{1}{2}\sum_k | g_{1k} \pm g_{2k} |^2/(E - \omega_k)$ and can be evaluated analytically \cite{Sanchez-Burillo2017,Gonzalez-Tudela2017,Lombardo2014} (see Appendix A).
Bound states both inside and outside the continuum appear as poles of Eq.~\eqref{eq:Amplitude}, that is, as zeros of
\begin{equation}
\label{eq:PoleEq}
F_{\pm}(E) = E - \varepsilon_\pm - \Sigma_{\pm}(E).
\end{equation}
The contribution of these bound modes to the dynamics can be obtained by evaluating the residues of Eq.~\eqref{eq:Amplitude} at the corresponding pole spectral positions \cite{John1990,Sajeev1994,gaveau1995}.

\begin{figure}
\includegraphics[width = \columnwidth]{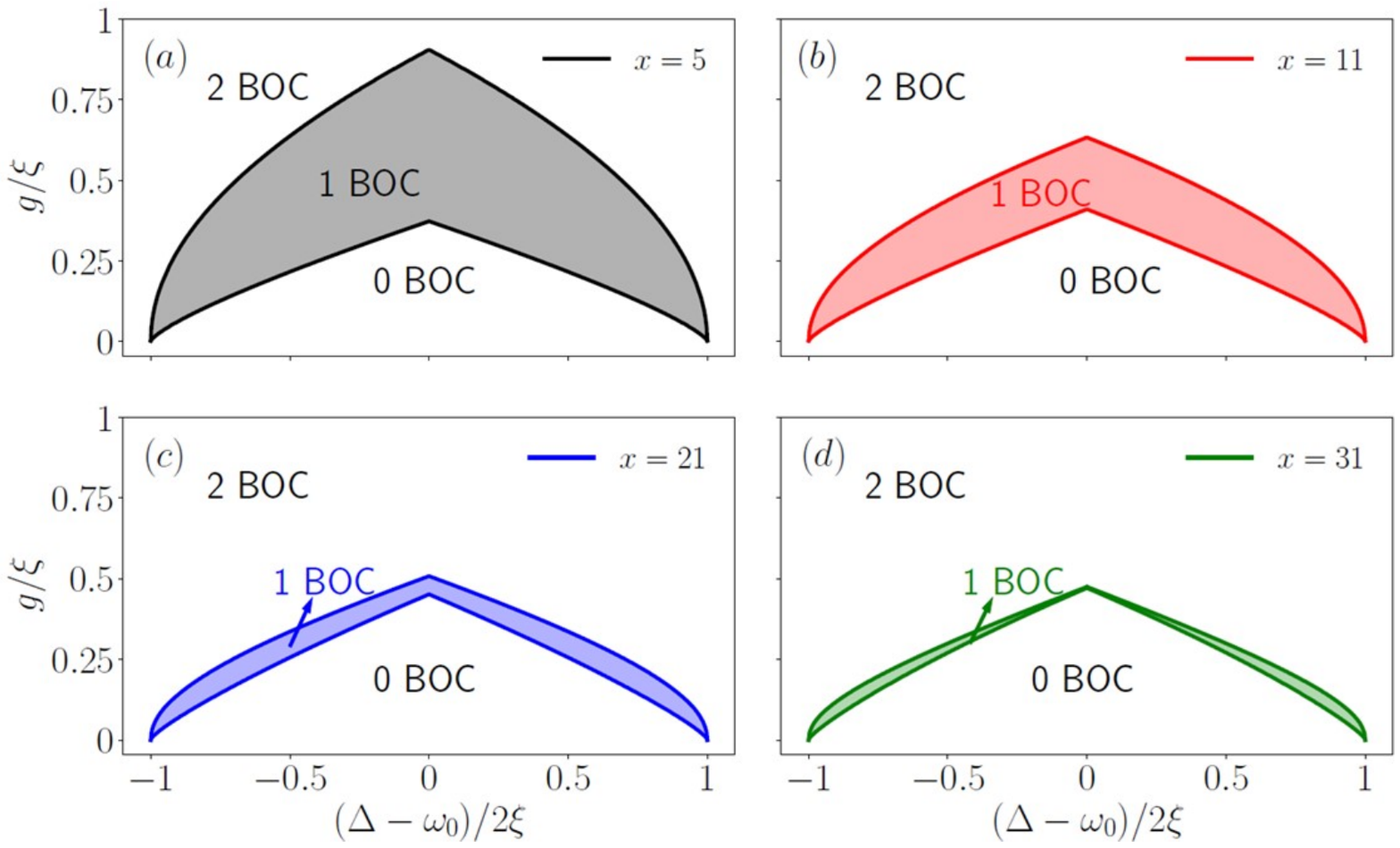}
\caption{
Regions of parameter space in which the system of two identical emitters coupled to a cavity-array waveguide (see Fig.~\ref{fig:Scheme}) supports zero, one, or two bound states outside the continuum. Each panel corresponds to a different inter-emitter separation: $x = 5$, $11$, $21$, and $31$ in panels (a)-(d), respectively.
The plots show the $\Delta$-$g$ parameter space, indicating the number of BOCs supported by the system. Below the solid curves no BOCs are present, while the regions between the curves, highlighted by the coloured background, correspond to the existence of a single BOC. Above both curves, the system supports two BOCs.
}
\label{fig:SpectraBoundModes}
\end{figure}
Figure~\ref{fig:SpectraBoundModes} illustrates, for several representative inter-emitter separations, the regions of parameter space in which the system supports zero, one or two BOCs.
In this figure, the search for BOCs is restricted to the case where the emitter excitation energy $\Delta$ lies within the photonic band, as this condition is necessary for the existence of BICs \cite{Gonzalez-Tudela2017}.
For all the inter-emitter distances considered, the qualitative behaviour is similar. Depending on the values of $\Delta$ and $g$, the system may support zero, one, or two BOCs. Notably, the region of parameter space supporting a single BOC progressively narrows as the inter-emitter separation increases.
Figure~\ref{fig:SpectraBoundModes} should be regarded as schematic, since the range of considered $g$s extends well beyond the regime of validity of the rotating-wave approximation, which is expected to break down for $g \gtrsim 0.1$ \cite{Niemczyk2010,Nori17}. In particular, BOCs do not appear when $\Delta$ lies near the centre of the band unless the coupling strength reaches values for which counter-rotating terms can no longer be neglected.
By contrast, when $\Delta$ is close to the band edges, BOCs may still arise for values of $g$ compatible with the rotating-wave approximation. 
In the following, we therefore focus on this latter coupling regime. Throughout, we fix the inter-emitter separation to x = 31 and investigate the conditions under which one or two BOCs coexist with a BIC.

\begin{figure*}
 \includegraphics[width = 2\columnwidth]{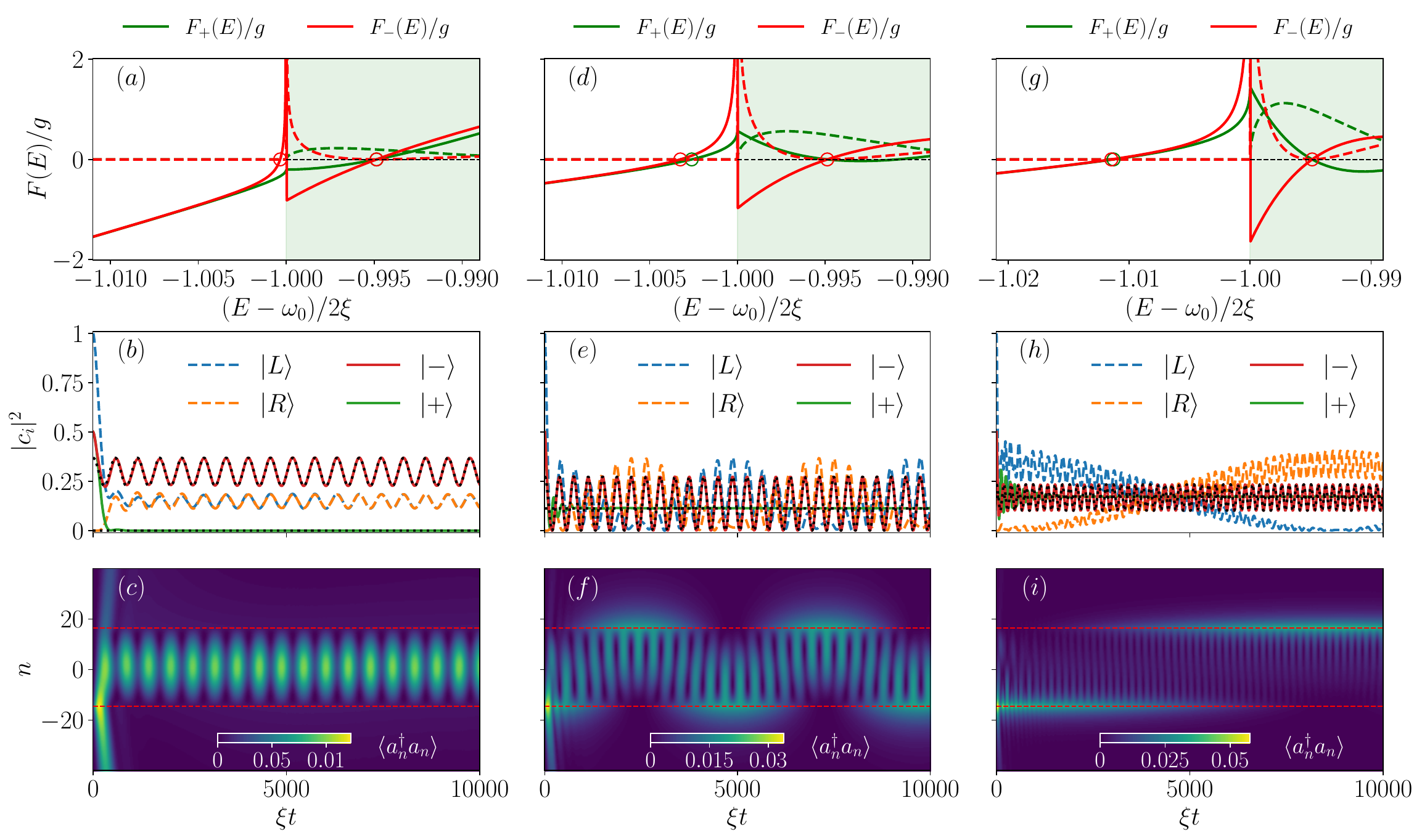}
 \caption{
Illustration of three spontaneous emission processes in which an initially excited emitter relaxes into different superpositions of bound states involving both the emitters and the waveguide field.
In all cases, the parameters are chosen such that the lowest-energy BIC participates in the dynamics: $\xi = 1$ and $\Delta = \omega_0 - 2\xi \cos[(x - 1)\pi/x]$, where $x = 31$ is the number of cavities separating the two emitters.
Each column corresponds to a different emitter-waveguide coupling strength: $g = 0.02$, $0.05$, and $0.1$ for the left, central, and right columns, respectively.
The left column [panels (a)-(c)] shows decay into a superposition of one BIC and one BOC, while the remaining columns correspond to decay into a superposition of one BIC and two BOCs.
Panels (a), (d), and (g) display the functions $F_{\pm}(E)$ for the symmetric (green curves) and antisymmetric (red curves) matter sectors. The real and imaginary parts of $F_{\pm}(E)$ are shown as solid and dashed lines, respectively. The solutions of Eq.~\ref{eq:PoleEq} are marked by circles, colour-coded according to the corresponding $\pm$ sector, and the shaded green region denotes the waveguide energy band.
Panels (b), (e), and (h) show the time evolution of the excited-state populations of the emitters. Populations in the $\ket{L(R)}$ basis are shown with dashed lines, while those in the $\ket{\pm}$ basis are shown with solid lines. The long-time asymptotic predictions, obtained from the bound-state contributions, are indicated by dotted black lines.
Finally, panels (c), (f), and (i) show the spatial distribution of the photon occupation number generated during the spontaneous emission process. The positions of the two emitters within the cavity array are marked by vertical red dashed lines.
}
 \label{fig:LargeFigure}
\end{figure*}
Figure~\ref{fig:LargeFigure} presents three distinct spontaneous emission processes in a WQED system, corresponding to different values of the emitter-waveguide coupling strength $g$. In each case, an initially excited leftmost emitter evolves into a superposition of bound states, both inside and outside the continuum. We fix the inter-emitter separation to $x = 31$ cavities and take the two emitters to be identical, with transition energies placed inside the photonic band but close to the lower band edge, $\Delta = \omega_0 - 2\xi \cos\!\left[(x - 1)\pi/x\right]$.
This choice gives rise to BOCs that are strongly localised around the emitters and therefore possess a large matter component, resulting in a pronounced influence on the spontaneous emission dynamics.
The organisation of Fig.~\ref{fig:LargeFigure} is as follows.  The top panels [(a), (d), and (g)] show the solutions of the bound-state equations for the symmetric and antisymmetric matter sectors, corresponding to the $\ket{+}$ and $\ket{-}$ combinations of emitter excitations. The middle panels [(b), (e), and (h)] display the time evolution of the excited-state populations of the left and right emitters, together with the contributions from the bound modes. The bottom panels [(c), (f), and (i)] present the spatiotemporal evolution of the photon occupation number along the waveguide.

The left column corresponds to the weak-coupling case $g = 0.02$. In this regime, Fig.~\ref{fig:LargeFigure}(a) shows that the equation $F_{-}(E) = 0$ admits two solutions, marked by circles: one lying within the continuum (shaded green region) and one outside it. By contrast, no solutions are found for $F_{+}(E) = 0$, indicating that all bound-state wavefunctions reside in the antisymmetric matter sector $\ket{-}$. For the BIC, this antisymmetric character is enforced by the specific choice of the cavity separation between the emitters.
For the BOCs, the single-emitter bound states that emerge close to the band edge couple through their evanescent photonic tails. This interaction leads to level repulsion, such that the antisymmetric combination $\ket{-}$ becomes more strongly bound, while the symmetric combination $\ket{+}$ is pushed into the band and turns into a resonance rather than a true bound state.  As a side remark, had the emitter transition frequency been chosen close to the \textit{upper} band edge, the situation would be reversed: the $\ket{+}$ combination would become more localised, while the $\ket{-}$ state would have evolved into a resonance.
Figure~\ref{fig:LargeFigure}(b) shows the time evolution of the excited-state populations, expressed both in the localised basis $\ket{L(R)}$ and in the symmetric and antisymmetric basis $\ket{\pm}$.
The occupation of the state $\ket{-}$ exhibits persistent oscillations in the steady state. After a transient regime, the populations are well described by the contributions from the poles of Eq.~\eqref{eq:Amplitude}, demonstrating that these oscillations arise from the coherent superposition of the two bound states within the same symmetry sector.  
As a consequence, the long-time dynamics is characterised by a single oscillation frequency, given by the energy difference between the two bound modes shown in Fig.~\ref{fig:LargeFigure}(a).
Such dynamics, in which decay into the waveguide is suppressed while coherent emitter-emitter exchange is preserved, constitute an example of a decoherence-free subspace for multi-emitter systems coupled to structured reservoirs. Related behavior has been demonstrated experimentally with two giant atoms in a braided configuration in a non-structured reservoir \cite{KannanNat2020}, indicating that bath-induced non-Markovianity can be harnessed to promote interference effects in multi-emitter waveguide QED.

Figure~\ref{fig:LargeFigure}(c) illustrates the dynamics of the emitted field, with the emitter positions indicated by red dashed lines. A significant fraction of the emitted light remains trapped between the emitters, forming a ``breathing'' mode localised around the central cavity.  
This photonic mode oscillates coherently with the populations of the $\ket{L(R)}$ states, providing clear evidence of a non-local, entangled oscillatory steady state. 
This  trapped oscillating mode formed between two spatially separated emitters acting as atomic mirrors effectively constitutes a tunable supercavity within the cavity array \cite{ZhouPRA2008, Zhang2023MagicCav}. 
%
Panels~\ref{fig:LargeFigure}(d), (e), and (f) correspond to a larger coupling strength, $g = 0.05$, chosen to explore the regime in which the system evolves into a superposition of three bound states: two outside the continuum and one inside it.
Figure~\ref{fig:LargeFigure}(d) shows the bound-state solutions for this configuration. The stronger emitter-waveguide coupling leads to increased confinement of the single-emitter BOCs, which consequently appear at energies further away from the band edge compared to the weak-coupling case shown in panel (a).  
This enhanced localisation reduces the effective coupling between the bound modes, resulting in a smaller energy splitting between the $\ket{\pm}$ symmetry sectors.  As a consequence, the $\ket{+}$ mode also becomes bound, in addition to the two bound modes associated with the $\ket{-}$ symmetry.
This configuration gives rise to the more complex dynamics displayed in panels~\ref{fig:LargeFigure}(e) and (f), where the steady state contains contributions from three oscillatory components. The resulting dynamics is governed by the energy differences between each pair of bound states.  
Notably, in this regime the populations in both the matter and photonic sectors oscillate between the left and right emitters, rather than exhibiting the breathing-mode pattern observed in the weak-coupling case.
This behaviour is a direct consequence of the presence of two BOCs with opposite parity. The resulting slow dynamics, governed by the small energy splitting between these modes, leads to an effective population exchange between the emitters, such that the $|L\rangle$ and $|R\rangle$ populations essentially swap in time.

Finally, the rightmost panels correspond to an even stronger coupling, $g = 0.1$. For this coupling strength and the chosen value of $\Delta$, the single-emitter BOCs become so strongly localised that their mutual coupling is significantly suppressed, leading to a small splitting between the BOCs with $\ket{\pm}$ symmetry.  
Aside from the different oscillation timescales arising from the distinct energy separations between the bound states, the qualitative time evolution of the populations closely resembles that observed in the intermediate-coupling regime shown in the central column.

\begin{figure}
    \centering
    \includegraphics[width=\columnwidth]{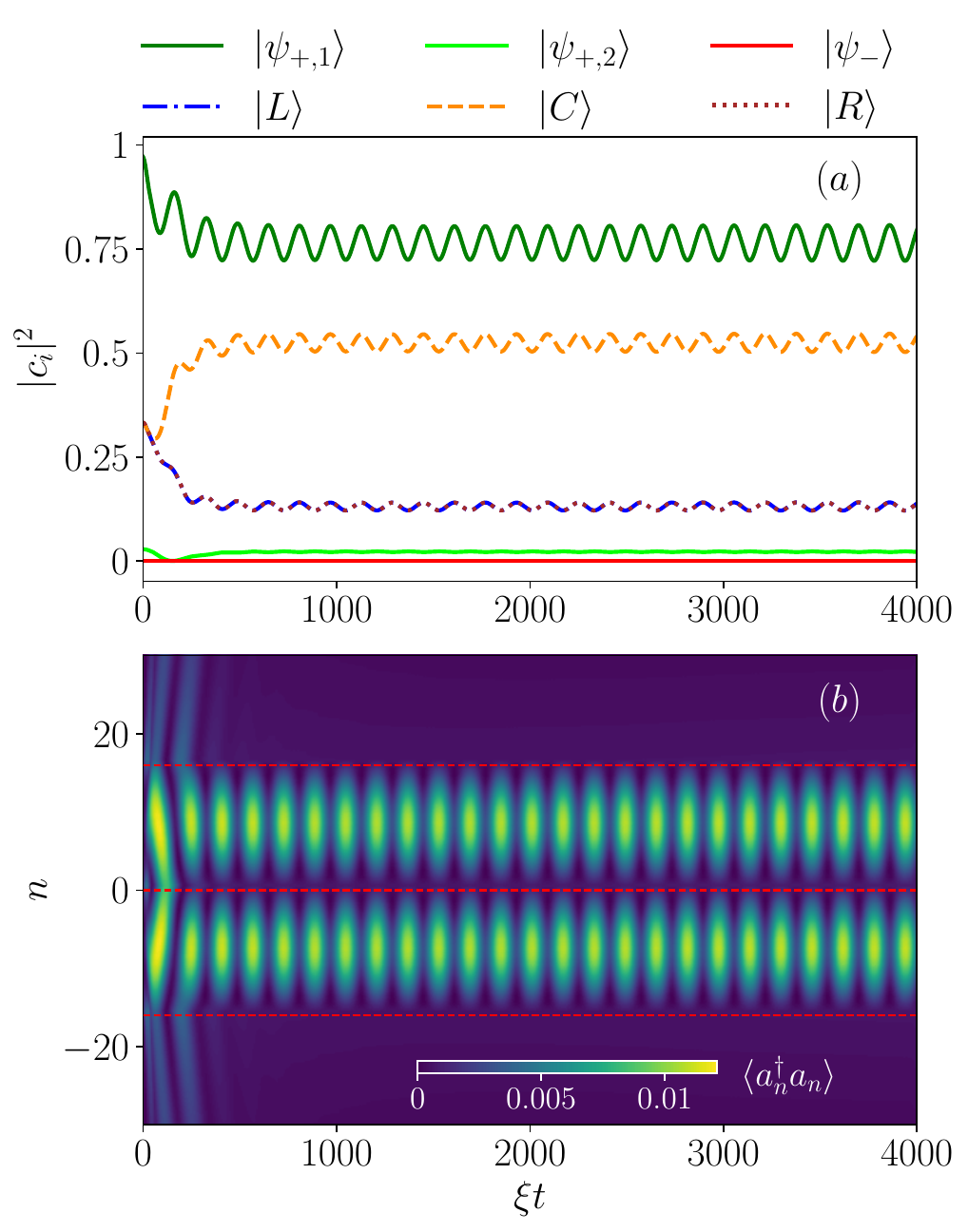}
    \caption{
Spontaneous emission leading to oscillatory dynamics between two states involving three emitters and the waveguide field.  
The emitters are located at positions $n = -16$, $0$, and $16$, corresponding to the left ($L$), central ($C$), and right ($R$) emitters, respectively, and are initially prepared in the fully symmetric state $ \ket{\psi}_{\mathrm{in}} = (\ket{L} + \ket{C} + \ket{R}) / \sqrt{3}$.
(a) Time evolution of the excited-state populations of the emitters.
The populations of the collective emitter states $\ket{\psi_{+,1}}$ (green), $\ket{\psi_{+,2}}$ (lime), and $\ket{\psi_{-}}$ (red) are shown as solid lines. 
The populations of the individual emitter states $\ket{L}$, $\ket{C}$, and $\ket{R}$ are shown as a blue dash-dotted line, an orange dashed line, and a brown dotted line, respectively.  
(b) Time evolution of the emitted waveguide field. The positions of the emitters are indicated by vertical red dashed lines.  
The emitter-waveguide coupling strength is $g = 0.02$, while the remaining parameters are the same as in Fig.~\ref{fig:LargeFigure}.}
    \label{fig:3atoms}
\end{figure}
We have also examined how the dynamics scale when more than two emitters are coupled to the waveguide. While the two-emitter case benefits from a basis of symmetric and antisymmetric excited states that remain dynamically decoupled, this property is specific to $N=2$. For $N>2$ no complete set of matter-only single-excitation states exists that remains mutually uncoupled under the Green function dynamics. Nevertheless, parity symmetry is preserved, allowing coherent dynamics to arise when the system is initialized in a state with well-defined parity.
Figure \ref{fig:3atoms} illustrates this behavior for three emitters located symmetrically along the waveguide at sites $n=-16$, $0$, and $16$, labeled $L$, $C$, and $R$, respectively. 
The system is initially prepared in the fully symmetric single-excitation state
$ \ket{\psi_{\mathrm{in}}} = \frac{1}{\sqrt{3}}\left(\ket{L} + \ket{C} + \ket{R}\right) $,
which has even parity and therefore selectively excites the even-parity collective modes of the system. Panel (a) shows the time evolution of the excited-state populations. After an initial decay, the system evolves into sustained coherent oscillations between two even-parity collective states, while the odd-parity mode remains unpopulated. Consequently, the excitation probabilities of the individual emitters oscillate in time, with the outer emitters exhibiting in-phase dynamics distinct from that of the central emitter.
Panel (b) displays the corresponding time evolution of the waveguide field. Following an initial emission burst, the field develops localized, time-periodic oscillations around the emitters, synchronized with the collective excitation dynamics. These spatiotemporal field patterns provide a direct manifestation of the coherent superposition of bound states governing the long-time behavior of the system.
Such coherent oscillatory dynamics are characteristic of small emitter arrays, particularly for $N=3$ and $N=4$. 
For larger $N$, the dynamics involve an increasing number of parity-defined collective states and become progressively more complex, requiring increasingly fine-tuned system parameters to sustain similar behavior.

%
To summarise, we have studied the existence and generation of superpositions of bound states, both inside and outside the continuum, in a cavity-array waveguide coupled to two spatially separated emitters. We have identified the parameter regimes required for the emergence of these states through spontaneous emission, analysed their dynamical formation, and provided analytical predictions for their long-time behaviour.

Under these conditions, a system initially prepared in a product state with a single excited emitter evolves, under free dynamics, into a non-local equilibrium state given by a superposition of bound modes. By suitably tuning the system parameters, we have shown that spontaneous emission can drive the system into an stationary regime in which both the emitter populations and the waveguide field exhibit persistent oscillations. Correspondingly, the photon density in the vicinity of the emitters undergoes coherent oscillations, evolving in synchrony with the excited-state populations of the emitters.

These superpositions of bound states, involving components both inside and outside the continuum, enable the generation of entanglement between distant emitters mediated by the waveguide, while simultaneously giving rise to a well-defined breathing mode in the photonic density of the intermediate cavities. We have also examined how these dynamics scale with the number of emitters. While the two-emitter configuration provides the most transparent and controllable analytical description, systems with three or more emitters generally exhibit increasingly complex collective dynamics, in which the photonic occupation displays richer spatial and temporal structures that obscure a simple breathing-mode interpretation, even though oscillatory behaviour between parity-defined collective states may still persist for small arrays.

It is worth noting that all waveguide QED platforms considered in this work are experimentally feasible using state-of-the-art superconducting circuit architectures, as well as photonic crystal waveguide implementations \cite{vanDiePen2025}, highlighting the practical relevance of our results.

\begin{acknowledgments}
 We acknowledge projects PID2023-148359NB-C21 and CEX2023-001286-S (financed by MI-CIU/AEI/10.13039/501100011033) and the Government of Aragon through Project Q-MAD. C.A.G-G acknowledges financial support from SECIHTI Mexico, under grant No. CBF2023-2024-2888, and DGAPA-PAPIIT-UNAM under grant No. IA104625.
\end{acknowledgments}

\appendix
\section{Appendix A: Analytical solutions for BICs for two distant emitters}
In this appendix we analytically determine the energies of bound states inside the continuum using Eq. \eqref{eq:PoleEq}.
We begin by computing the self-energies within the photonic band, $E \in [\omega_0-2\xi,\omega_0+2\xi]$.
\begin{align}
\label{eq:SelfEnComp}
\Sigma_{\pm}(E+i0^{+}) &= \frac{1}{2}\sum_k \frac{\lvert g_{1k}\pm g_{2k} \rvert^2}{E-i0^{+}-\omega_k} \nonumber \\
&=\frac{g^2}{N_k} \sum_k \frac{1\pm \left(e^{ikx} + e^{-ikx}\right)}{E-i0^{+}-\omega_k} \nonumber \\
&=\frac{ig^2}{2\sqrt{4\xi^2-(E-\omega_0)^2}} \left(1 \pm e^{ik(E)x} \right),
\end{align}
where the wave vector $k(E)$ is defined through $\cos[k(E)] = (E-\omega_0)/2\xi$.
Introducing the spectral density of a two-level system $J(E) = g^2/\sqrt{4\xi^2-(E-\omega_0)^2}$, Eq. \eqref{eq:SelfEnComp} can be rewritten as
\begin{equation}
\label{eq:SigmaJw}
\Sigma_{\pm}(E) = i\frac{J(E)}{2} \left(1\pm e^{ik(E)x}\right).
\end{equation}
The self-energy in Eq. \eqref{eq:SigmaJw} vanishes when $k(E) = m\pi /x$, where $m$ is an odd integer for the $\ket{+}$ sector and an even integer for the $\ket{-}$ sector.  
As a result, the bound-state condition $F_{\pm}(E) = 0$ in Eq.~\eqref{eq:PoleEq} is satisfied for energies $E = \Delta = \omega_0 - 2\xi \cos( m \pi x)$. 
In the cases illustrated in Fig. \ref{fig:LargeFigure} we choose $m=x-1$ which corresponds to the BIC with an energy closest to the lower band edge.
\bibliography{ref}
\end{document}